# Light-induced emergent phenomena in 2D materials and topological materials

Changhua Bao[1], Peizhe Tang[2,3,*], Dong Sun[4,*] & Shuyun Zhou[1,5,*]

[1]*State Key Laboratory of Low-Dimensional Quantum Physics and Department of Physics, Tsinghua University, Beijing 100084, P. R. China.*

[2]*School of Materials Science and Engineering, Beihang University, Beijing 100191, China.*

[3]*Max Planck Institute for the Structure and Dynamics of Matter, Center for Free Electron Laser Science, 22761 Hamburg, Germany.*

[4]*International Center of Quantum Matter, Peking University, Beijing 100871, P. R. China.*

[5]*Frontier Science Center for Quantum Information, Beijing 100084, P. R. China.*

[*] Email: peizhet@buaa.edu.cn, sundong@pku.edu.cn, syzhou@mail.tsinghua.edu.cn

Abstract

Light-matter interaction in 2D and topological materials provides a fascinating control knob for inducing emergent, non-equilibrium properties and achieving new functionalities in the ultrafast time scale (from fs to ps). Over the past decade, intriguing light-induced phenomena, e.g., Bloch-Floquet states and photo-induced phase transitions, have been reported experimentally, but many still await experimental realization. In this Review, we discuss recent progress on the light-induced phenomena, in which the light field could act as a time-periodic field to drive Floquet states, induce structural and topological phase transitions in quantum materials, couple with spin and various pseudospins, and induce nonlinear optical responses that are affected by the geometric phase. Perspectives on the opportunities of proposed light-induced phenomena as well as open experimental challenges are also discussed.

## Introduction

The past decade has witnessed rapid progress in our understanding of quantum materials, particularly in the fields of quasi-2D materials and topological materials thanks to the advanced sample preparation techniques, more systematic understandings of topological physics, and the development of various advanced ultrafast time-resolved measurements. From the material point of view, atomically thin materials[1], assembled van der Waals heterostructures[2], together with twisted bilayer structures[3-5] provide attractive playgrounds for creating new structures with tailored properties. Extension of topological physics from insulators to semimetals has led to the discovery of topological semimetals[6,7] such as Weyl, Dirac, nodal-line (NL) semimetals and unconventional topological semimetals with higher topological charges[8-10]. Moreover, instrumentation developments in time- and angle-resolved photoemission spectroscopy (TrARPES), time-resolved X-ray diffraction, ultrafast electron diffraction (UED) and time-resolved transport measurements have also provided unprecedented opportunities to capture the formation of new steady states, dynamics of lattices and transport properties in non-equilibrium.

The research fields of 2D materials and topological materials are not independent, but rather they are highly interconnected, with graphene acting as a bridge. The Haldane model, in which the quantum anomalous Hall (QAH) effect was predicted as a topological phenomenon, is based on the graphene lattice[11]. From the theoretical point of view, graphene is at the topological phase boundary between a trivial insulator and a 2D topological insulator (TI) when the system is driven by a strong spin-orbit coupling (SOC)[12]. Moreover, other layered or quasi-2D materials, such as bulk $MoTe_2$[13,14], $WTe_2$[15] and $PtTe_2$[16] as well as monolayer $WTe_2$[17] have also been found to be topological materials. Finally, the light–matter interaction can change the topological properties of the host materials, thereby further bridging the physics of 2D materials and topological materials.

Light–matter interaction plays critical roles, not only as an experimental probe for equilibrium properties, but more importantly as a control knob for inducing emerging non-equilibrium properties that are otherwise not possible in the equilibrium state. According to the light-matter interaction mechanisms, light-induced emerging phenomena can be roughly classified into four classes as schematically shown in Fig.1. First, the material properties can be dynamically manipulated in the ultrafast time scale through Floquet engineering[18,19], where light acts as a time-periodic electric field (Fig. 1a). Second, light can perturb the energy landscapes by injecting energy to electrons and lattices[20-25](Fig. 1b), leading to photo-induced phase transitions or excitations. Third, by coupling the angular momentum of light with the electron spin and various pseudospins[26-30] with different optical selection rules (Fig. 1c), detection and manipulation of spin and various pseudospins can be achieved. Fourth, the sensitivity of light-matter interaction to the geometric phase of the Bloch wave functions (Fig. 1d) can lead to exotic nonlinear optical responses in topological materials and 2D materials[31-34]. The light-matter interaction is used as a control knob for tailoring material properties dynamically for the first two classes, while for the latter two classes, it is used more as an experimental probe of the material properties with potential application in light-matter interaction manipulations. We note that the above simplified classification is not absolute and light-induced phenomena could involve multiple mechanisms. For example, the topological phase transitions can be induced not only by Floquet engineering but also by the dynamic tuning of the energy landscape via nonlinear phononic coupling and electron-phonon interactions, as have been shown for $ZrTe_5$[25,35], $MoTe_2$[24,36] and $WTe_2$[36,37].

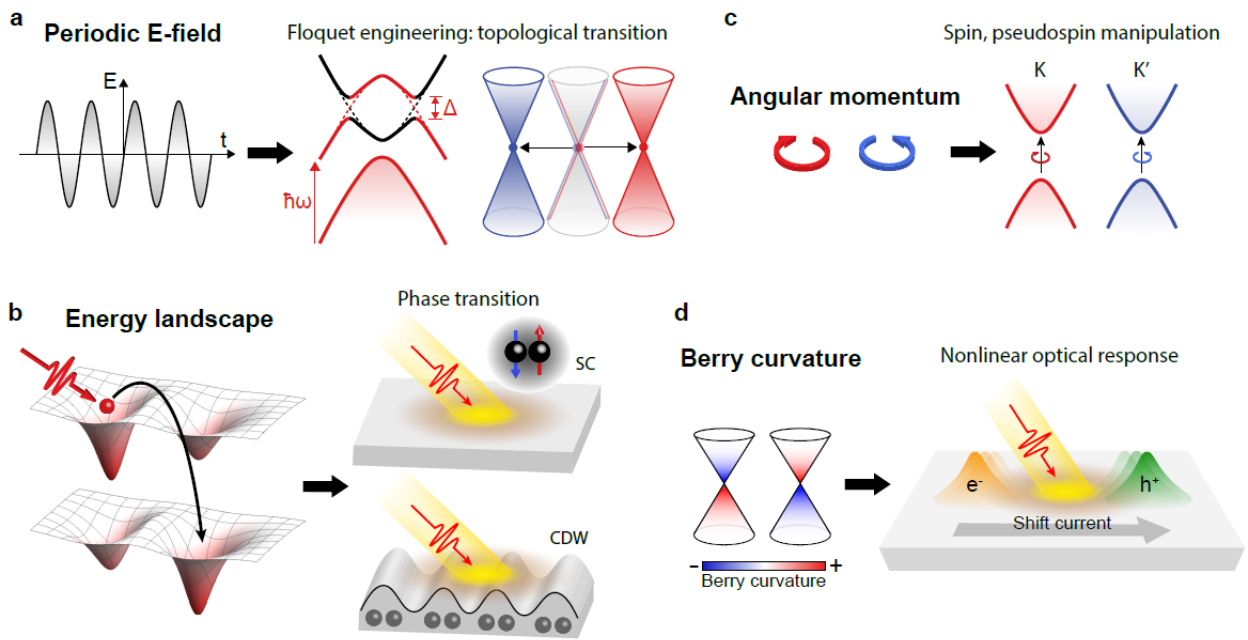

**Figure 1: Coupling of light with 2D and/or topological materials and light-induced emerging phenomena**. **a |** Floquet sidebands, band gap opening, band inversion and topological phase transition through Floquet engineering by using light as a time-periodic electric field. **b |** Energy carried by light can perturb the energy landscape, leading to phase transitions, such as light-induced superconductivity (SC) and charge density wave (CDW). **c |** Coupling of light to spin and pseudospin through optical selection rules using different light polarizations. **d |** The interplay between light and geometric phase contributing to exotic quantum effects such as nonlinear optical response.

While major progress has been achieved in light-induced phenomena, so far this field is still at a rapidly developing stage, with many predicted phenomena awaiting experimental realization. A deep understanding of light-induced phenomena is not only important for revealing the fundamental physics, but also useful for device applications since many quantum devices are based on the interaction with light[38-42]. Moreover, achieving dynamic tuning of the material properties in the picosecond (ps) to femtosecond (fs) time scale is critical for device applications operating at an ultra-high speed. In this Review, we provide a summary on the light-induced phenomena discovered so far, outline intriguing theoretical proposals and identify experimental challenges to overcome. The field of emergent light-induced phenomena has been quite active in the past decade, and we look forward to more exciting discoveries with joint experimental and theorical efforts.

Key points:

1. Two-dimensional materials and topological materials have evolved as important research frontiers for condensed matter physics and materials sciences. Light-matter interaction plays critical roles in emerging exotic phenomena in 2D materials and topological materials not only as an experimental probe, but also as a control knob for inducing emergent non-equilibrium properties that are otherwise not possible to be achieved in the equilibrium state.

2. Light, regarded as a time-periodic electric field, can induce the photo-dressing Floquet states, which can be further utilized to dynamically engineer the electronic properties of quantum materials especially topological properties, dubbed as the Floquet engineering.

3. By resonantly exciting electrons or lattices, light-matter interaction can dynamically change the energy landscape of 2D and topological materials, leading to the light-induced phase transitions, e.g., the emergency of light-induced superconductivity or hidden states.

4. By coupling the angular momentum of light with spins and pseudospins, light-matter interaction can be used to detect and manipulate various quantum degrees of freedom for new concepts of device applications.

5. By coupling to geometric phase of the Bloch wave functions, light-matter interaction can be used as powerful probes of geometric phase related properties, and to manipulate the material's response, leading to rich nonlinear optical responses.

# Floquet engineering

Light can act as a time-periodic electric field to couple with electrons in solid-state materials and lead to the formation of quasi-energy sidebands through multi-photon absorption or emission, which are called the Floquet states in analogy to the Bloch states (Fig. 2a). The strong light-matter coupling could change the symmetry of the host materials and result in dynamic engineering of the electronic structures, including the band gap opening and the band inversion that are critical for determining the topological properties in non-equilibrium. Such Floquet engineering can be used as a dynamic control knob for tailoring the properties of topological materials or dynamically inducing a topological phase transition.

# Floquet states

In solid state physics, the spatially periodic Hamiltonian $H(\vec{r}) = H(\vec{r}+\vec{R})$ results in Bloch states[43] whose energy bands have the discrete translational symmetry in the momentum space, $E(\vec{k}) = E(\vec{k} + n\vec{G})$ ($n = \pm 1, \pm 2, ...$), where $\vec{k}$ is the momentum and $\vec{G}$ is the reciprocal lattice vector. In analogy, by applying a time-periodic driving field to the solid-state system, it is possible to form a steady Floquet state and the related Hamiltonian has the discrete time-translational symmetry $H(\vec{k},t) = H(\vec{k},t+T)$. In this case, the time-averaged physical quantity over the period $T$ of the driving laser field does not change. Therefore, in the Floquet theory, the time-dependent Schrödinger equation in the Hilbert space $\mathcal{H}$ can be solved by mapping it to a time-independent infinite eigenvalue problem in an extended Hilbert space $\mathcal{H} \otimes L_T$, where $L_T$ represents the space of "multi-photon" dressed states. The solution of time-dependent Schrödinger equation $\psi_l(k,t)$ has the form of $\psi_l(k,t) = e^{-i\overline{\varepsilon_l(k)}t/\hbar}\phi_l(k,t)$, where $e^{-i\overline{\varepsilon_l(k)}t/\hbar}$ is a periodic phase factor and $\hbar$ is the reduced Planck constant. $\phi_l(k,t)$ could be expanded by discrete Fourier series, $\phi_l(k,t) = \sum_m e^{im\omega t}u_l^m(k), m \in Z$, where the photon frequency is $\omega = 2\pi/T$. By solving the time independent equation under the Floquet basis $\sum_n H_{mn}(k) u_l^m(k) = [\varepsilon_l(k) + m\hbar\omega]u_l^m(k)$, we have the time-averaged Floquet energy $\overline{\varepsilon_l(k)} = \frac{1}{T}\int_0^T \varepsilon_l(t,k)dt = \varepsilon_l(k) + m\hbar\omega$, where $\varepsilon_l(k)$ is the eigenvalue of the Floquet Hamiltonian, and its matrix element has the form of $H_{mn}(k) = \frac{1}{T}\int_0^T dt H(k,t)e^{i(m-n)\omega t}$ (Fig. 2b). Herein we would like to emphasize that states of $\phi_l(k,t)$ and $e^{im\omega t}\phi_l(k,t)$ with the eigenvalues of $\varepsilon_l(k)$ and $\varepsilon_l(k) + m\hbar\omega$ represent the same Floquet state. Therefore, we could define the first Floquet Brillouin Zone (BZ) in $(-\frac{\hbar\omega}{2}, \frac{\hbar\omega}{2})$, and those states with eigenvalues beyond the first Floquet BZ can be obtained through the multi-photon absorption or emission from states inside the first Floquet BZ[44].

For Bloch states, hybridization with the folded bands at the BZ boundary (in momentum) can lead to a band gap opening. Similarly, in the Floquet theory, the electronic properties of the host materials could be changed by coupling with optical field. For example, a gap can open for 2D Dirac materials at the band crossing[19] and at the first Floquet BZ boundary[19,45-51]. This application of a time-periodic driving field can be used as a control knob for engineering non-equilibrium material properties, dubbed as Floquet engineering. It is important to note that the Floquet theory shown above does not provide any information about the occupation of the Floquet states, which is extremely important for non-equilibrium steady state and is determined by the competition between the laser driving and the relaxation of excited quasiparticles in an open system[52]. Thus, although the Floquet theory provides a clear understanding of the light-matter interaction in the strong coupling limit, it is based on a simplified approximation and does not consider electron occupation and relaxation.

Floquet-Bloch states were theoretically proposed in graphene, where application of a circular polarized light (CPL) is predicted to lead to a gap opening at the Dirac point (Fig. 2c, d) and a photo-induced Hall current [19]. Unlike the conventional Hall effect which is induced by a static magnetic field in equilibrium, the photo-induced Hall current is caused by strong light-matter interaction which breaks the time-reversal symmetry. The topological properties of such non-equilibrium, periodically driven systems are further analyzed[47], and an analogy to the Haldane model[11] is discussed. A nonzero Chern number has been obtained, and topologically protected edge states and quantum Hall effect have been proposed[50,53-57]. Floquet engineering has also been extended to other Dirac fermion systems, such as the surface state of TIs[50]. In addition to hexagonal lattice like graphene, it has been suggested that topological states with helical edge states can also be induced in a topologically trivial semiconductor quantum well[58,59].

Following these early proposals, there has been a surge of theoretical efforts on the Floquet engineering of quantum materials. For review articles on the theoretical development of Floquet engineering, we refer the readers to these papers[18,60-66]. There are fewer experimental examples of Floquet states, which are made available by advanced time-resolved techniques, such as TrARPES[67] and time-resolved transport measurements by pumping under extreme experimental conditions. This is at least partly attributed to the stringent experimental conditions such as low photon energy with an extremely high electric field strength required for the excitation laser, as well as the unresolved problems regarding the occupation, scattering and dissipation mechanisms in such an open system under real experimental conditions. Joint experimental and theoretical efforts are needed to tackle these critical questions along the experimental realization of Floquet engineering.

Experimental examples of Floquet states

To realize Floquet states, a low excitation photon energy in the mid-infrared (MIR) range with an intensive light field in the range of $10^7$-$10^8$ V/m is required[19], which can only be satisfied by using ultrashort laser pulses with tens to hundreds of fs pulse duration. Direct experimental evidence of Floquet states has been demonstrated through TrARPES measurements of a TI material $Bi_2Se_3$ with 120 meV[68] and 160 meV[69] MIR excitation and an electric field strength of 2.5-3.3×$10^7$ V/m applied on the sample surface. The choice of MIR pump has three major considerations. Firstly, this is below the intrinsic bulk band gap of $Bi_2Se_3$ in equilibrium, which is useful to suppress direct optical transitions from the bulk states and the resulting sample heating. Secondly, such suppression allows the application of a larger electric field (E) and leads to a stronger intensity for the higher order sidebands (m≥2) whose intensity scales as $E^{2m}$ under a weak field approximation[69]. Thirdly, the dynamical gap at the avoided crossing point scales inversely with the pump photon energy and a lower photon energy could lead to a detectable gap[19].

Experimentally, Floquet sidebands of the topological surface states have been observed in $Bi_2Se_3$ at energies displaced by $m\hbar\omega$ in the ultrafast time scale when the pump overlaps coherently with the probe pulse[68,69] (Fig. 2e). There are two experimental signatures to indicate Floquet engineering. One is the observation of a dynamical gap generation of 62 meV at the crossing points between the topological surface state and sideband replicas along the momentum direction perpendicular to the electric field for both linear (p and s polarizations) and circular polarized lights (CPL)[68]. The other is the observation of a gap of 53 meV at the Dirac point of the surface state due to the breaking of time-reversal symmetry by CPL[68]. In addition, by switching the pump polarization between p and s, selective excitations of Floquet states versus Volkov states have been demonstrated, which correspond to photon-dressed initial electronic states (inside the sample) and photon-dressed free-electron-like final states (after photo-excited from the sample into the vacuum) respectively[69].

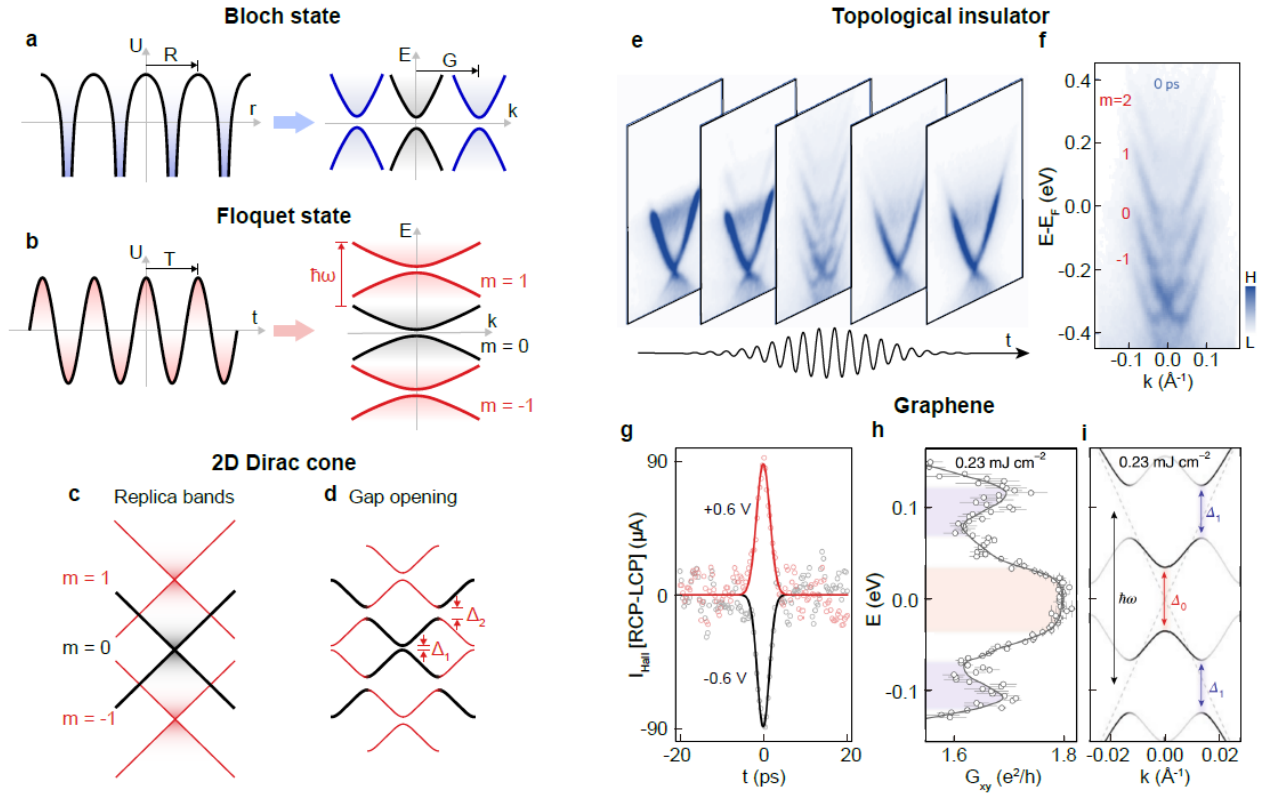

**Figure 2: Floquet engineering and experimental evidence of Floquet states. a |** Spatially-periodic potential and Bloch bands in the k-space. **b |** Time-periodic potential and Floquet bands in energy. **c,d |** Floquet engineering in 2D Dirac system leading to Floquet sidebands (red) and resonant gap opening at the crossing point. **e,f |** Experimental observations of Floquet states by TrARPES in TI $Bi_2Se_3$. TrARPES spectrum of surface Dirac cone at different delay times (e). TrARPES spectrum at the zero delay time (f). **g |** Light-induced anomalous Hall current signals. **h |** Light-induced Hall conductance as a function of energy. **i |** Effective band structure under light excitation using the Floquet theory. Panel e are TrARPES data from REF.[69] and reprinted from REF.[291], Springer Nature Limited. Panel f reprinted from REF.[69], Springer Nature Limited. Panels g-i reprinted from REF.[71], Springer Nature Limited.

Although graphene has been proposed as a model system for realizing topological Floquet states since 2009[19,50,55,57,70], signatures of Floquet states in graphene have been demonstrated only recently by light-induced anomalous Hall effect (AHE) or photovoltaic Hall effect[71]. CPL excitation at photon energy of 191 meV with a field strength of $4.0 \times 10^7$ V/m (corresponding to a pump fluence of 0.23

mJ/cm[2]) is used to generate a dynamical gap in graphene, and anomalous Hall currents are reported to be possible characteristics of Floquet states according to the following two experimental features. First, the anomalous Hall current is transverse to the applied longitudinal current and it reverses polarity for opposite light helicities (Fig. 2g). Second, a conductance plateau is observed with a width of 60 meV when the Fermi energy is tuned inside the gap (Fig. 2h), suggesting a gap opening consistent with the calculated light-induced gap of 69 meV (Fig. 2i). However, by taking the relaxations of photo-excited carriers into account, theoretical simulations point out that although Floquet states are formed in the strong light-matter coupling limit, the experimentally observed AHE in graphene is not only contributed by the topologically non-trivial Berry curvature of the Floquet states, but also by the population imbalance in these dressed bands[72-74]. The experimental observation of AHE is enabled by the development of intensive MIR laser and the application of ultrafast time-resolved transport measurements using a laser-triggered ultrafast photoconductive switch[75], which extends the time resolution of transport measurements to the picosecond time scale. The significance of this experimental work is therefore not only in reporting the AHE in graphene, but also it opens new opportunities for detection of light-induced AHE[50] and light-induced superconductivity[22] using ultrafast time-resolved transport measurements. The gap opening in graphene inferred by time-resolved transport measurements of such Floquet engineering could in principle be detectable by TrARPES, which still awaits experimental progress.

## Tailoring topological materials

Dirac and Weyl semimetals are 3D analogues of graphene[6,7,76], whose topological properties have been widely investigated in the equilibrium state. Floquet engineering can be applied to these materials to achieve highly tunable topological properties in non-equilibrium[77] (Fig. 3a). Floquet engineering can be used to dynamically generate a Weyl semimetal from a ground state TI by renormalizing the bulk energy gap of a TI[78]. Starting with a Weyl semimetal as ground state, Weyl nodes with opposite chiralities acting as magnetic monopoles in the momentum space can be tuned by Floquet engineering. Applying CPL shifts the position of the Weyl nodes, and a net AHE in the plane orthogonal to the incident beam is predicted without considering the non-equilibrium charge distribution[79,80]. For Dirac semimetal as ground state, CPL could break the time-reversal symmetry and split Dirac fermions into a pair of Weyl fermions (WFs)[80-82], and their separation is tunable by light intensity or direction[82] and the WF can be even tilted to type-II WF with highly tilted Weyl cones[81]. Such light-induced Weyl semimetal hosts a chiral pumping effect[80], which originates from a quantum anomaly to induce the charge pumping from the surrounding reservoir and has the contribution to the transport similar to the chiral magnetic effect in equilibrium Weyl fermions.

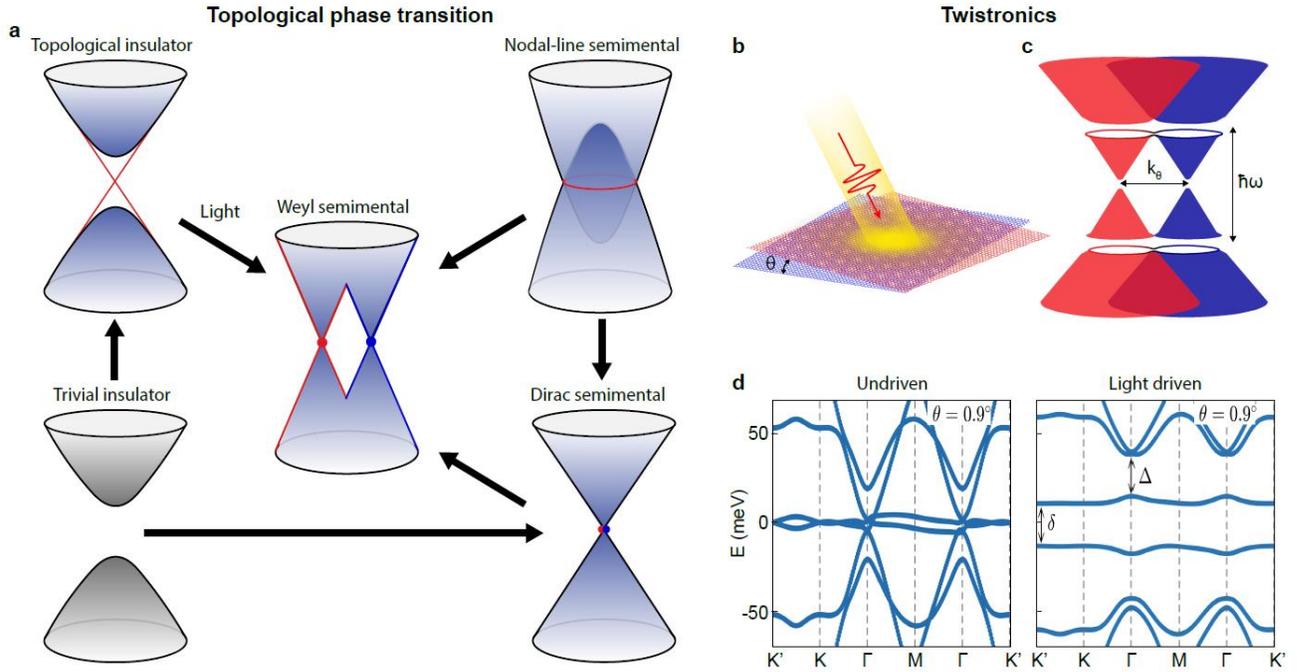

**Figure 3: Predicted light-induced topological phase transitions by Floquet engineering. a |** A schematic summary of light-induced topological phase transitions between topologically non-trivial and trivial materials. **b |** Interplay between light and twisted materials with moiré pattern. **c |** Interaction between the two Dirac cones in twisted bilayer graphene and light-induced band flatting and gap opening. **d|** Calculated band structure of twisted bilayer graphene with and without light. Panels d reprinted from REF.[112], CC BY 3.0 (https://creativecommons.org/licenses/by/3.0).

In addition to Dirac and Weyl semimetals, nodal-line semimetals[83] with band-crossing lines provide another rich playground for Floquet engineering. CPL excitation of nodal-line semimetals can lead to Weyl semimetals with a larger monopole charge[84,85] and the emergence of an anomalous light-tunable Hall conductivity[81,86-88]. Furthermore, a light-induced topological phase transition from quantum spin Hall (QSH) to QAH has been proposed in monolayer FeSe[89], in which the superconductivity survives at high temperature in equilibrium[90] with high-order topology[91-93]. QSH has been reported in FeSe[94] and CPL is further predicted to induce an effective SOC to lift the spin degeneracy that can lead to QAH.

### Inducing topological properties

Perhaps the most fascinating aspect of Floquet engineering is that it can even induce a dynamic topological phase transition in topologically trivial materials. This idea was proposed in a HgTe/CdTe semiconductor quantum well, where light-induced pseudospin reconfiguration is expected to result in non-trivial topology[58]. Gap opening and band inversion can in principle be induced through Floquet engineering, making it possible to turn a trivial insulator into a TI without the need to change its structural properties. This concept has been applied to a wide range of band insulators[95-101], such as black phosphorous[102], silicene[103] and transition metal dichalcogenides[104]. In black phosphorous, the light-induced topological phases are theoretically shown to be particularly rich, including Dirac semimetal, QAH phase[105], type-II and a novel type-III Dirac semimetal[102]. They are also highly tunable by changing the direction, intensity and frequency of the excitation light.

### Moiré–Floquet engineering.

Floquet engineering has been recently extended to twisted layered materials. In magic-angle twisted bilayer graphene (MATBG) (Fig. 3b and 3c) where the twist angle has been used as a control knob to generate a flat band near the Fermi energy with a high density of states[4,106], emergent phenomena such as superconductivity[107] and strong correlation physics[108] have been observed experimentally without involving photo-excitation. Combining twistronics[3,4,109] with Floquet engineering can further lead to an even richer field of "opto-twistronics" or "moiré–Floquet engineering", where light excitations with different driving frequency, field strength and optically generated spatial period can be used to engineer flat bands, topological properties and valley selective excitations (Fig 3b)[110].

Floquet engineering can enhance the flat band feature of MATBG by increasing its flatness or tuning the magic twist angle. The flatness of the bands near the charge neutrality point and their separation from high energy bands can be improved by UV[111] or visible to near infrared (near-IR) light excitation[112,113] through modulation of the interlayer hopping and lattice relaxation (Fig. 3d). Furthermore, the magic twist angle can be tuned to be either larger or smaller by changing the frequency of light[114], thereby increasing the range of magic twist angle for the flat band. Moreover, in analogy to the moiré superlattice potential, a spatially-periodic potential with different symmetry and period generated by laser fields can also be used to induce flat bands. Along this line, flat bands induced by light-field generated Kagome lattice have been proposed by using MIR light excitation[115].

Floquet engineering of the Berry curvature using CPL can also lead to topological phase transitions in twisted materials[116]. The flat bands can carry nonzero Chern numbers when the sample is irradiated by CPL to break the time-reversal symmetry[111]. Light-induced valley-polarized Floquet Chern flat bands and tunable large Chern numbers have also been predicted in twisted multilayer graphene[117]. In twisted transition metal dichalcogenide bilayers, longitudinal light in a waveguide is expected to lead to a topological phase transition by tuning the interlayer coupling strength[118].

Light control of the valley and layer pseudospins can also be included by extending Floquet engineering to twisted multilayer graphene, for example, strong optical drives can be used to realize valley-selective quasi-energy gaps in twisted double bilayer graphene[119]. Light-induced valley-dependent layer polarization could also be an important pathway for topological superconductivity in bilayer transition metal dichalcogenides[120,121]. Compared to light field in free space, longitudinal light generated from a waveguide can have a more dramatic effect to renormalize the interlayer tunneling and the Fermi velocity of the quasi-energy spectra of MATBG[114].

The application of Floquet engineering to twistronics is a rather new science frontier, yet it is an active field with many intriguing proposals in the last two years (see review article[110] for theoretical proposals). Moiré-Floquet engineering provides an attractive pathway to flat band engineering, however, there are some additional experimental challenges along their experimental realization using TrARPES measurements due to the difficulties in the detection of flat bands, which exist only within a small energy and momentum region[122,123] and call for higher energy and momentum resolution.

## Experimental challenges

While theoretical progress in Floquet engineering has evolved rapidly in the past decade, there has been less experimental progress, with data reported only on $Bi_2Se_3$[68,69], graphene[71] and recently, $WSe_2^{124}$. Although TrARPES can directly measure the renormalized band structure from Floquet engineering, there exist several fundamental experimental challenges.

Table 1: Summary of photon wavelength and light power used for Floquet engineering in different materials.

| Material | Gap size at DP (meV) | Wavelength (μm) | Light power (V/m) | Reference |
|---|---|---|---|---|
| Graphene | 200 | 0.8 | $10^9 \sim 10^{10}$ | REF[57] |
| Graphene | 40 | 1.5 | $2 \times 10^8$ | REF[19] |
| Graphene | 60 | 6.5 | $4 \times 10^7$ | REF[71] (**exp**) |
| Graphene | 6 | 43 | $3 \times 10^5$ | REF[45] |
| $Bi_2Se_3$ | 53 | 10 | $2.5 \times 10^7$ | REF[68] (**exp**) |
| Material | TPT | Wavelength (μm) | Light power (V/m) | Reference |
| FeSe | QSH to QAH | 0.16 | $10^9$ | REF[89] |
| Graphene | DP merge | 0.2 | $3.4 \times 10^{10}$ | REF[290] |
| TI | TI to Weyl | 0.4 | $10^9$ | REF[78] |
| Black phosphorus | NL to Dirac | 2.5 | $4 \times 10^8$ | REF[102] |
| $Na_3Bi$ | Dirac to Weyl | 0.8 | $5 \times 10^8$ | REF[82] |

Abbreviation: DP: Dirac point; TPT: topological phase transition; QSH: quantum spin Hall; QAH: quantum anomalous Hall; TI: topological insulator; Weyl: Weyl semimetal; NL: nodal-line semimetal; Dirac: Dirac semimetal.

An extremely strong light field in the range of $10^7$-$10^8$ V/m (Table 1) is typically required for Floquet engineering, which is quite close to the upper limit of the light field that 2D and topological materials can tolerate. An important limiting factor is the sample damage threshold, for example, the damage threshold at pump wavelength of 800 nm with 100 fs pulse duration is approximately $5 \times 10^8$ V/m for topological materials $Bi_2Se_3$ and black phosphorous according to our experimental experience. Another limiting factor is imposed by the multi-photon ionization, which generates a huge number of scattered photoelectrons that could wash out the signal of the electronic band structures.

Both the damage threshold[125,126] and the multi-photon ionization threshold[127] can be significantly increased by using a lower pump photon energy $\hbar\omega$. Moreover, the observable quantity, the gap at the Dirac point $\Delta_{DP}$ scales with the electric field $E$ and the pump photon energy $\hbar\omega$ by $\Delta_{DP} = \sqrt{4V^2 + \hbar^2\omega^2} - \hbar\omega \approx 2e^2 v_F^2 E^2/(\hbar\omega)^3$,[19] where V is $E/\omega$ with $E$ being the field strength and $v_F$ is the Fermi velocity. In order to have a detectable gap on the order of 10 meV, a lower pump photon energy in the MIR is favored. However, the photon energy cannot be too low, for example THz pump[128], for the observation of Floquet states, because of two reasons. First, THz light corresponds to an energy scale of meV which is smaller than the energy broadening of the dispersion measured by TrARPES (tens of meV)[129], and therefore the separation between Floquet sidebands (equal to the photon energy of only a few meV) is unlikely to be resolved. Second, strong THz light field could excite free carriers in solids and induce intraband electron scattering which may destroy the Floquet states[124]. Considering all these factors, a photon energy in the MIR range is preferred for realizing Floquet engineering.

In addition to the field strength and photon energy of the pump pulse for generating the Floquet states, another challenge is the UV probe light for detecting the transient band structures in TrARPES. For 2D materials such as graphene and transition metal dichalcogenides, the electrons near the Fermi energy are at momentum valleys with a large momentum value (1.7 Å$^{-1}$ for graphene). Therefore, high harmonic generation [130] using gas as nonlinear optical medium is required for generating the large probe photon energy to access the large momentum range. However, the high harmonic generation generally has a lower probe photon flux and undermines the energy resolution as compared to 5.9-6.2 eV probe light source generated by fourth harmonic generation[131]. This makes it difficult to resolve the small band gap induced by Floquet engineering. In addition, a tunable probe photon energy is also required to access the 3D Dirac and Weyl points at the right $k_z$ value.

Implementing TrARPES with versatile pump and probe pulses, for example, a strong pump pulse extending to the MIR range and a tunable probe pulse with a large and tunable photon energy to cover a large momentum range while maintaining the high energy- and momentum-resolution, is critical for advancing the experimental progress in light-induced phenomena through Floquet engineering.

## Light-induced phase transitions.

Light can induce phase transitions by perturbing the energy landscape (Fig. 4a). The energy carried by light can be absorbed resonantly and transferred via nonlinear couplings to other degrees of freedom of the materials, such as phonon and electron, leading to the change of the energy landscape in the excited state and thus resulting in the formation of the non-equilibrium state. Such interaction could lead to light-induced phase transitions, such as light-induced superconductivity, light-induced hidden or metastable states.

Strong light–matter interaction can change the energy landscape in the non-equilibrium excited state and induce a phase transition in the host materials beyond the Floquet scenario, for example insulator-metal transition in manganite[147-149], $VO_2$[145,150-153] and $TaS_2$[20,138,139,146,154]. Such photo-induced phase transition can be revealed by the carrier dynamics through time-resolved optical methods[132-137] and TrARPES measurements[20,131,138-141], and lattice dynamics through ultrafast X-ray diffraction[142-144] and ultrafast electron diffraction[137,145,146]. In contrast to the Floquet engineering discussed in the previous session, there is hardly any self-contained theory to explain all phenomena of light-induced phase transitions in 2D materials and topological materials, since an optical field with photon energy ranging from several meV to several eV could couple with many different kinds of degrees of freedom in quantum materials. Therefore, the underlying mechanism is case by case and strongly dependent on the electronic properties of specific materials and the pump photon energy.

Among possible scenarios to utilize the light field to change the energy landscape of 2D materials and topological materials, non-linear phononic coupling is regarded as one of efficient ways to achieve this goal. Laser fields in the MIR or THz range could perturb the lattice of solid-state materials through non-linear phononic coupling. Upon application of a light field with field strength as large as $10^7$ V/m, IR-active optical phonon modes $Q_{IR}$ with a small momentum could be excited directly, behaving like an oscillator under the driving force in the classical mechanical description. For systems with strong nonlinear phononic coupling, such as the ferroelectric, multiferroic and thermoelectric materials, the strongly dynamic driving of IR mode $Q_{IR}$ will change the motion of the other coupled phonon modes (like the Raman mode, $Q_R$) via nonlinear coupling $H_{nonl} = a_{12}Q^1_R Q^2_{IR} + a_{21}Q^2_R Q^1_{IR} + \ldots$. Here, $a_{12}$ and $a_{21}$ are the nonlinear coupling coefficients constrained by the crystal symmetry of host materials. Through this mechanism, the local minimum position of the oscillation of the coupled phonon mode can be shifted, just like applying an inner strain to the host materials. Such dynamic tuning can result in the formation of a metastable state in non-equilibrium, with the emergence of a wide range of exotic physical properties, including light-induced metal-insulator transition[155], superconductivity[21,142], ferroelectrics[156,157], magnetism[158,159] and topological phase transitions[160].

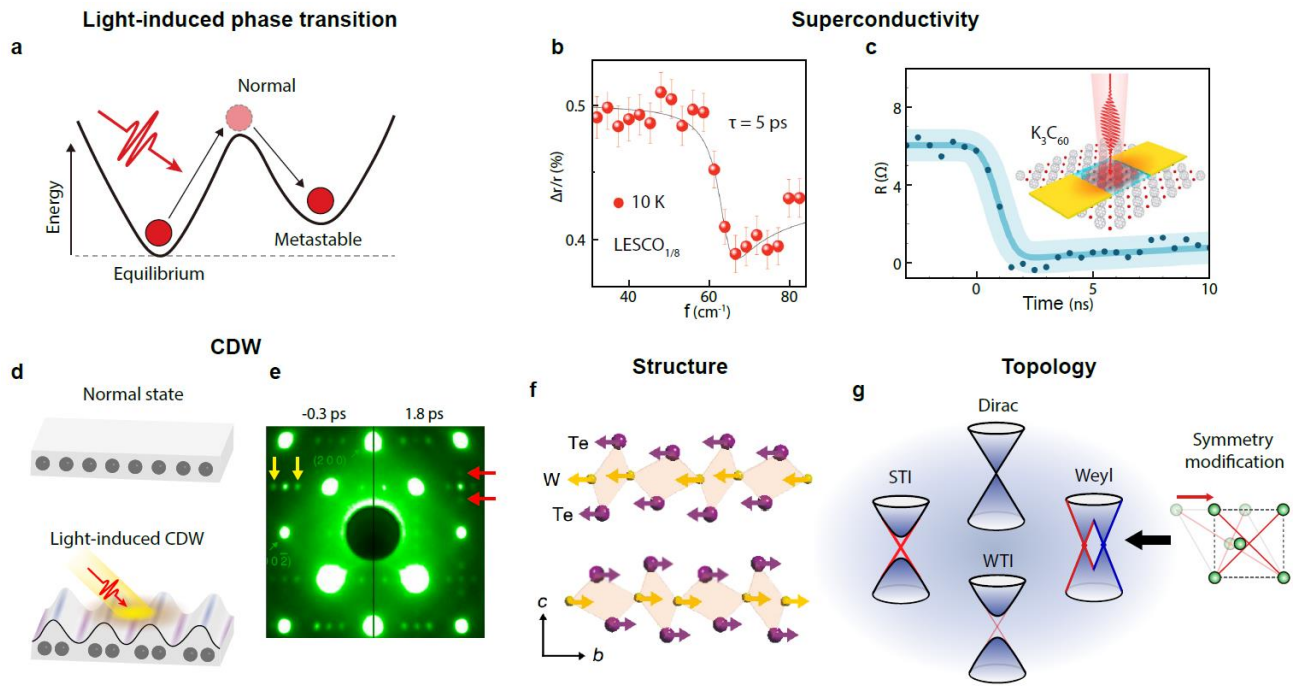

**Figure 4: Light-induced phase transitions**. **a |** Light-induced metastable state from equilibrium state leading to a phase transition. **b |** Transient c-axis reflectance of LESCO$_{1/8}$, normalized to the static reflectance. The appearance of a plasma edge at 60 cm$^{-1}$ demonstrates that the photoinduced state is superconducting. **c |** Resistance of a laser-irradiated K$_3$C$_{60}$ as a function of time after photoexcitation. **d |** Schematic of light-induced CDW. The light field can change the normal state to CDW state (light-induced CDW). **e |** Electron diffraction patterns before (left) and 1.8 ps after (right) photoexcitation with a near-infrared laser pulse. **f |** A light-induced interlayer shear phonon mode changes the inversion symmetry of a Weyl semimetal WTe$_2$. **g |** The light-matter interaction changes the symmetry of the host materials and leads to topological phase transitions. Panel b reprinted from REF.[21], AAAS. Panels c reprinted from REF.[162], Springer Nature Limited. Panel e reprinted from REF.[205], Springer Nature Limited. Panel f reprinted from REF.[24], Springer Nature Limited.

Superconductivity.

The most intriguing aspect of light–driven phase transitions is probably the light–induced superconductivity by mode-selective excitation[22]. This was demonstrated in the cuprate, $La_{1.675}Eu_{0.2}Sr_{0.125}CuO_4$ ($LESCO_{1/8}$), where resonant pumping at 80 meV (15 µm) suppresses the stripe phase at 1/8 anomaly and leads to the emergence of Josephson plasma resonance in the c-axis optical properties, indicating light-induced superconductivity[21] (Fig. 4b). Resonant excitation of $B_{1u}$ mode in $YBa_2Cu_3O_{6.5}$ at similar photon energy also enhances the inter-bilayer coherence and superconductivity as revealed by time-resolved optical measurements[161], and the underlying mechanism is attributed to large-amplitude apical oxygen distortion as resolved by time-resolved X-ray diffraction[142]. For $K_3C_{60}$, light-induced superconductivity has also been reported by electrical transport measurements[162] (Fig. 4c) in addition to optical measurements[163] at high temperature under the excitation of phonon modes. In contrast to the possible nonlinear phononic coupling shown in $YBa_2Cu_3O_{6.5}$, the observed amplitude of superconductivity in $K_3C_{60}$ as indicated by the reduction of spectral weight in the real part of optical conductivity, only depends on the integrated pulse area and is independent of the electric field strength[162]. Possible mechanisms behind the non-equilibrium superconducting state have been discussed[164-171], where the light-induced excitations are argued to enhance Cooper pairing[172-180], while the exact mechanism remains to be nailed down.

We note that in addition to light-induced superconductivity, photoexcitation has also been used to investigate the intrinsic properties of the superconducting state by weak perturbation. These studies have probed the dynamics of carrier diffusion in the superconducting and pseudogap states through time-resolved optical measurements[132-137] and Cooper pair evolution by TrARPES[131,140,141,181,182]. In addition, light-matter interaction could also induce exotic phenomena when the superconductors are excited by the THz laser field, for example, the resonance between the optical field and Higgs amplitude mode[168,178,183-191], the excitation of hidden superfluid stripe[192], the emergence of a Leggett mode associated with relative phase fluctuations between two superconducting order parameters in multiband superconductor[193,194] and gapless superconductivity with forbidden Anderson pseudo-spin precessions[195,196].

## Hidden or metastable phases

In 2D materials, phase transitions involving symmetry breaking, such as charge density wave (CDW)[197] and superconductivity[198], can occur. In most situations, there is a competition between different symmetry breaking states with a small energy difference, and finally the system falls into the most stable state with minimum energy in equilibrium. Other metastable states are well separated by the energy barrier and thus are hidden in equilibrium. In such systems, the light field can be used as a control knob to perturb or change the energy landscapes, thereby switching the system among symmetry breaking metastable states by exciting the system to overcome the energy barrier. 2D transition metal dichalcogenides[199] represent an ideal platform to explore the light-induced phase transition because their structures are sensitive to light, for example, light-induced permanent transition from 2H to 1T phase in $MoTe_2$[200] and transient transition from 2H to 1T phase in $MoS_2$[201] are already observed experimentally.

CDW systems are characterized by breaking specific translational symmetry. While light excitation usually destroys CDW order leading to CDW melting[20,138,139,146], it has been shown recently that light can also induce new CDW orders[154,202-206] (Fig. 4d). For example, a hidden metastable phase can be induced by a short laser pulse with a large drop in resistance in 1T-$TaS_2$[154]. Such hidden phase is switchable by subsequent laser pulse, electrical current, or heating, which makes it promising for future electronics applications. Very recently, a novel inverted CDW order was induced by light in 1T-$TaSe_2$[206]. The inverted CDW order shows high-metallicity in contrast to a typical CDW insulator. In addition to modification of the original CDW order, distinct CDW order can also be induced by light on an ultrafast time scale. In $LaTe_3$, a new CDW along the a-axis is induced by light in contrast to the original CDW along the c-axis[205] (Fig. 4e). In this work, light-induced topological defects[207,208] are proposed to play an important role in inhibiting the backward path, leading to a lifetime as long as 5 ps. Moreover, light can be used to manipulate the electronic dimensionality in 3D CDW material $TiSe_2$, which has been recently demonstrated by combining TrARPES and UED measurements[209].

## Topological phase transitions

Light-matter interaction can also modify the crystal symmetry, providing another pathway to induce a topological phase transition in addition to Floquet engineering. For example, the inversion symmetry plays a central role in determining the topological nature of materials[6] and light-induced inversion symmetry restoration has been demonstrated. For example, light can drive the shear mode phonon directly and induce a structural phase transition from $T_d$ to 1T' phase in WTe$_2$[24,36,37] and MoTe$_2$[23,36] (Fig. 4f). Meanwhile, the structure of the 1T' phase breaks the inversion symmetry and hosts type-II WFs; while the $T_d$ phase is centrosymmetric and topologically trivial[13,210-212]. Therefore, there is a light-induced topological phase transition from type-II Weyl semimetal to a trivial metal in WTe$_2$. Such light-induced inversion symmetry restoration leads to a periodic reduction of the spin splitting that has been observed by TrARPES[213]. It has been proposed that applying an electrical current is another effective way to break the inversion symmetry, which can be probed by second harmonic generation. Therefore, a current-induced second harmonic generation (SHG) effect is expected in Dirac or Weyl semimetals[214]. In addition, we note that more crystalline symmetries besides the inversion symmetry can also be broken by photocurrent in Weyl semimetal TaAs, leading to possible Weyl node manipulation[215].

Light-matter interaction has also been applied to tune the topological properties of layered topological material ZrTe$_5$, which is close to the boundary between strong TI and weak TI and is therefore very sensitive to external perturbations. Coherent excitation of A$_{1g}$ Raman mode at 1.2 THz (5 meV) leads to a transition from a strong TI to a metastable weak TI[25]. A light-induced inversion symmetry breaking is also reported by exciting the IR-active phonon mode B$_{1u}$ using laser wavelength of 800 nm (1.5 eV) to drive ZrTe$_5$ from a strong TI to a Weyl semimetal[35]. Such argument is supported by the emergence of a circular photogalvanic effect (CPGE) with giant dissipationless photocurrent in ZrTe$_5$.

## Coupling to spin and pseudospin

2D materials and topological materials exhibit intriguing physics involving spin and various pseudospins, such as sublattice, valley and chirality pseudospins. The degrees of freedom of light, including energy, polarization, angular momentum, can be tuned to be selectively coupled to spin or certain pseudospin following specific optical selection rules. Precise excitation and control of spin and pseudospins can be achieved by controlling the light while employing the strong coupling with spin and various pseudospins at the same time.

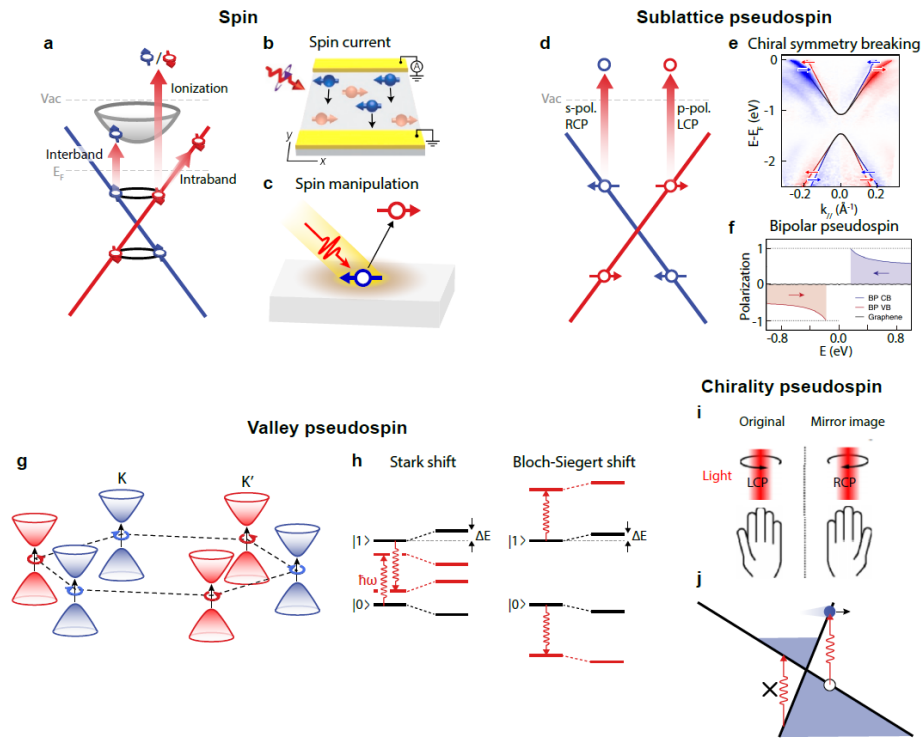

**Figure 5: Emerging properties due to coupling to spin, sublattice pseudospin, valley and chirality. a |** Three types of polarization-dependent electronic excitations: interband excitation, intraband excitation and ionization. **b |** A schematic of the light–induced spin current. Only the electrons with left spin move down as indicated by black arrows. **c |** Spin manipulation by different light polarizations. **d |** Polarization selective pseudospin excitations. **e |** Chirality-resolved dispersion to reveal the chiral symmetry breaking near the Dirac point of a Kekulé-ordered graphene. **f |** Opposite pseudospins in the valence and conduction bands of black phosphorus. **g |** Valley selective optical excitation by CPL. **h |** Schematics of the optical Stark shift and Siegert shift. **i |** Schematic of the chirality of light. **j |** Schematic to show that optical excitation in one branch is blocked by Pauli blockade in a tilted Dirac cone. Panel b reprinted from REF.[26], Springer Nature Limited. Panel e reprinted from REF.[237], CC BY 3.0 (https://creativecommons.org/licenses/by/3.0). Panel f reprinted from REF.[28], Springer Nature Limited. Panel h reprinted from REF.[29], Springer Nature Limited. Panel i reprinted from REF.[30], Springer Nature Limited.

## Spin

TIs are characterized by a strong SOC with spin-momentum locking for the topological surface state[216], and spin texture that has been resolved by spin-ARPES[217]. The light polarization can be used to detect such helical spin texture in TIs[218-222] and chirality related spin in Weyl semimetals[223-225] through polarization-dependent ARPES and to further manipulate the spin[26,27,226]. Left and right circular polarized (LCP and RCP) light can selectively excite electrons with spin up and spin down in the Dirac cone of TI surface state (Fig. 5a). Because the spin is locked with the momentum in TI surface states, such selective excitation potentially leads to spin-polarized photo-current, and the current and spin direction can be switched by the handedness of light[26,227] (Fig. 5b). Such spin-polarized photo-current switch can be extended to an ultrafast time scale (hundreds of fs) by low-energy THz light excitation, where sub-cycle spin-polarized photo-current is directly observed by TrARPES thanks to the long mean free paths of Dirac electrons on the surface states[128]. In addition, the coupling of light with spin of the photoelectrons can also lead to selective spin flipping by using different polarized lights. For example, linear polarized light flips the spin along the polarization direction, while RCP and LCP turn the spin into the in-plane and out-of-plane directions respectively[27,226] (Fig. 5c).

## Pseudospins

In 2D and topological materials, various pseudospins build up rich quantum degrees of freedom that can be coupled to and thus be controllable by light. The concept of pseudospins have been firstly applied to describe the two sublattice degrees of freedom in graphene[228], and later extended to the two momentum valleys of graphene[229] and transition metal dichalcogenides[230] (valley pseudospin), the two layers in a bilayer materials[231] (layer pseudospin) and the two relative directions, either parallel or antiparallel, between momentum and spin direction of WFs (chirality pseudospin)[232].

The most commonly discussed pseudospin is the sublattice pseudospin in the hexagonal lattice of graphene[228] and black phosphorous[28,233]. The pseudospin of photoelectrons can be resolved through polarization-dependent ARPES measurements by coupling to different light polarizations[234-236] (Fig. 5d). This unique coupling between light polarization and pseudospin has been recently applied to visualize chiral symmetry breaking near the Dirac point (Fig. 5e) of a Kekulé-ordered graphene[237]. Similar light-pseudospin coupling exists in black phosphorous which possesses similar yet puckered honeycomb lattice, and a novel bipolar pseudospin semiconductor is revealed based on such light-pseudospin coupling[28,238] (Fig. 5f).

In 2D hexagonal transition metal dichalcogenides such as $MoS_2$, $MoSe_2$, $WS_2$ and $WSe_2$, valley pseudospin exhibits rich coupling with light. Light control of valley pseudospin is mainly based on a valley-dependent selection rule for CPL as a result of the valley contrasting orbital angular moment:

the interband transition at K (K′) couples only to LCP $\sigma^+$ (RCP $\sigma^-$) (Fig. 5g)[230]. A linearly polarized photon, which is a coherent superposition of LCP and RCP photons, can transfer optical coherence into excitonic valley coherence of the host materials. Experimentally, this is verified on WSe$_2$ where the linear polarization angle of photoluminescence (PL) from neutral excitons always coincides with the excitation polarization[239]. Using circular polarized femtosecond pulses, the exciton level in each valley can be selectively tuned by as much as 18 meV through valley-selective optical Stark effect, equivalent to an ultrafast and ultrahigh valley pseudo-magnetic field[38,240]. The optical Stark effect can be understood as the repulsion of a pair of Floquet states between the original states (with energy $E_0$) under off-resonance light excitation, while the repulsion of another pair of Floquet states outside the original states can lead to Bloch-Siegert shift (Fig. 5h). Since Bloch-Siegert shift is proportional to $1/(E_0+\hbar\omega)$ as compared to $1/(E_0-\hbar\omega)$ of Stark shift, a Bloch-Siegert shift ($\approx$ 10 meV) can be entirely separated from the optical stark shift observed in WS$_2$ using CPL with a large detuning ($|E_0-\hbar\omega|\gg0$) Furthermore, because these two effects obey opposite selection rules at different valleys, the Bloch-Siegert shift and optical Stark shift can be selectively confined to one valley or the other by controlling the light helicity[29].

The chirality of the WF or multifold topological fermion, which is defined by whether the directions of spin and motion are parallel or anti-parallel[232], is another typical type of pseudospin. It shares similar circular optical selection rules with the valley degree of freedom as described in the previous paragraph, and CPL excites opposite sides of the WFs of opposite chiralities. The selection rule is a consequence of the angular momentum conservation during a spin-flip vertical transitions on the Weyl point. Due to the coupled spin and momentum for a crossing with a fixed chirality, it can generate a current under circular polarized excitation in a single tilted Weyl point, producing an experimentally observable CPGE (Fig. 5i,j). However, experimental observation of this CPGE relies on tilted Weyl cones with well-selected transition photon energy and suitable Fermi level as demonstrated in TaAs[241], otherwise the sum of photocurrent from a Weyl node pair must vanish identically. Furthermore, CPL can optically induce chirality that leads to gyrotropically ordered phase in certain materials with achiral phase. This has been realized experimentally in 1T-TiSe$_2$ by shining MIR CPL while cooling it below the charge density wave transition temperature. The chirality of this state is confirmed by the direction of circular photogalvanic current which depends on the optical induction[30].

## Nonlinear optical response

In topological materials, the effect of the geometric phase of the Bloch wave functions can have a profound impact on the optical response[242], especially the nonlinear optical response, through light-matter interaction. Such light-matter interaction can be used as powerful probes of geometric phase related properties, which are difficult to detect using other experimental probes. Alternatively, the interaction can be used to manipulate the material's response, leading to rich nonlinear optical response features in topological materials.

An early example of a nonlinear response caused by geometric phase is that even orders of high harmonic generation from monolayer $MoS_2$ crystal are compatible with the anomalous transverse intraband current arising from the material's Berry curvature[243]. More recent experimental works have focused on the nonlinear optical response of a diverging Berry curvature, which corresponds to a magnetic monopole in the momentum space of Weyl semimetals or multifold topological fermion materials[8-10,244] (Fig. 6a). The nonlinear optical response has been understood by nonlinear susceptibilities involving the geometric phase of the wave function[245,246]. By further applying a Floquet formalism described by Floquet bands of electrons dressed by photons, it has been shown that various nonlinear optical effects, including second order nonlinear responses such as shift current, SHG and third order nonlinear optical responses such as nonlinear optical Hall effect and Kerr rotations, are related to either the Berry connection (for second order response) or the Berry curvature (for third order response) of the electronic band in equilibrium[31]. This theoretical treatment is quite general and applicable to any solid, including Weyl semimetals and multifold topological fermion materials. Topological materials are ideal material platforms to observe such effects experimentally, because these nonlinear responses can be enhanced by the diverging Berry connections or curvature at the vicinity of the topological nodes in Weyl semimetals or multifold fermions.

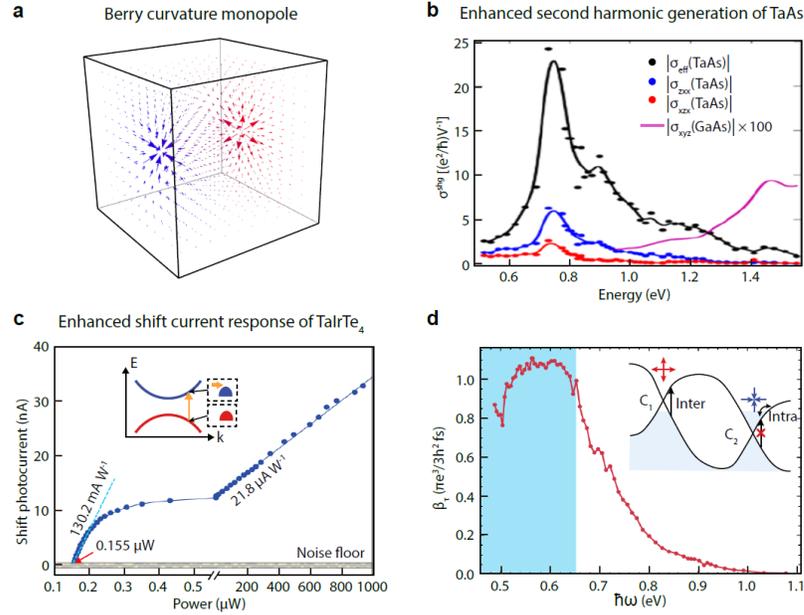

**Figure 6: Geometric Phase Effect on Nonlinear Optical Response. a** | Vector plot of the Berry curvature of Weyl semimetals in the momentum space. **b** | Nonlinear optical conductivity components $|\sigma_{zxx}|$, $|\sigma_{xzx}|$ and $|\sigma_{eff}| = |\sigma_{zzz} + 4\sigma_{xzx} + 2\sigma_{zxx}|$ of type-I Weyl semimetal TaAs as a function of incident photon energy through second harmonic generation measurement. The nonlinear optical conductivity component $|\sigma_{xyz}|$ of GaAs is multiplied by 100 and plotted for comparison. **c** | Light-power dependence of the photocurrent at 4 μm incident wavelength in a FET based on TaIrTe$_4$, a type-II Weyl semimetal. The inset illustrates the shift current generation after interband excitation. **d** | Circular photogalvanic effect of RhSi. CPGE amplitude as a function of photon energy, showing abrupt quenching above 0.65 eV. The inset illustrates intraband versus interband effects in Weyl semimetals. When inversion and mirror symmetries are broken, Weyl nodes of opposite chiralities are generated at different energies. Panel **b** adapted from REF.[32], APS. Panels **c** adapted from REF.[41], Springer Nature Limited. Panel **d** adapted from REF. [258], AAAS.

The major experimental challenge in identifying topologically enhanced nonlinear responses is to provide clear evidence that the enhancement is indeed due to topological features of host materials. Such measurements usually require the optical transition to be at the vicinity of the topological node. The earliest experimental study in this regard is SHG, which was performed on TaAs, a well-established type-I Weyl semimetal. The measurement at 1.5-eV photon energy already provides a SHG coefficient of d = 3600 pmV$^{-1}$ at 1.5 eV, 100 times larger than that of typical nonlinear medium GaAs[247], but the transition is quite far away from the topological node, which weakens the connection between this giant response and low energy WFs[127]. Further photon energy dependent measurement down to 0.56 eV shows the optical response is maximized at 0.7 eV (Fig. 6b)[32]. The position of the peak wavelength at 0.7 eV is complicated by optical response involving van Hove singularities when compared with the linear optical-conductivity spectrum, and the relatively large SHG response at 1.5 eV corresponds to the high-energy tail of a resonance at 0.7 eV.

Compared to SHG, the giant shift current responses observed in Weyl semimetals are more clearly confirmed to be related to a geometric phase effect by two parallel experiments on type-I TaAs and type-II TaIrTe$_4$ respectively[33,34]. The photocurrent measurement on TaAs at excitation photon energy of 117 meV, which is quite close to the Weyl node, shows a glass coefficient of G = 1.65×10$^{-7}$ cm V$^{-1}$, several orders of magnitude larger than that of other typical nonlinear media[33]. Parallel work on TaIrTe$_4$ demonstrates a very high responsivity of 130 mA/W in the low-power region at 4 $\mu$m and its relation to the Berry connection is further verified by comparing wavelength dependent measurements and numerical simulations[34] (Fig. 6c). Because the shift current is predicted to be proportional to 1/$\omega$, in type-II Weyl semimetals[248], it scales up for low energy photon. Thus the enhanced shift current response by the geometric phase opens an interesting possibility of using this category of materials for long wavelength detection applications[41].

More recently, studies of third order nonlinear optical effects have also been reported on WTe$_2$. Optical Kerr rotation measurements have indicated a third order nonlinear Kerr rotation arising from the Berry curvature hexapole allowed by the crystal symmetries[249], but it remains to be investigated whether the observed effects are related to the low energy WFs. More studies on higher order nonlinear responses of these category of topological semimetals await future explorations. Recent theoretical works have also been extended to antiferromagnetic (AFM) topological materials, such as AFM noncollinear Weyl semimetal and AFM TI, to further take into account the role of time-reversal symmetry[250-254]. The AFM TI MnBi$_2$Te$_4$ and AFM Weyl semimetal Mn$_3$Sn are the two model materials for such studies. Similar geometric phase enhancement on the magnetic counterpart of shift/injection current and harmonic generation are predicted for both of these, but experimental verifications are still quite limited[255].

Another example of geometric phase effect on the nonlinear optical response is the quantization of the injection current tensor associated with optical transitions near a Weyl node. The injection current generation rate (dj$_i$/dt), which can lead to experimentally measurable CPGE, is theoretically demonstrated to be proportional to the Berry monopole charge of a single Weyl cone in the non-interacting limit, which is independent of the frequency of the excitation light over certain wavelength range and the specific properties of material[256] (Fig. 6d). To observe this geometric phase effect experimentally, a Weyl material without the mirror symmetry is required, as the nodes of opposite charges are degenerate in energy, which leads to exact cancellation of the CPGE response from a pair of perfectly symmetric Weyl nodes. For this consideration, multifold fermion materials, which also support Chern bands with a monopole structure at the nodal point, provide an ideal experimental platform to observe the quantization effect[10,257] (Fig. 6d). Experimentally, terahertz emission from transient generation of circular polarization dependent photocurrent are measured. The measurement with excitation photon energy range of 0.48-1.1 eV on RhSi shows a plateau switch around 0.65 eV, a photon energy that is consistent with a quantized injection current generation interpretation. The CPGE contribution comes from the four-fold fermion at the $\Gamma$ point only when the excitation photon energy is below 0.65 eV, whereas the transitions at the six-fold fermion R points are Pauli blocked[258]. However, a parallel measurement with excitation photon energy from 0.2-1.1 eV on CoSi shows a peak instead of a plateau switch at 0.4 eV[259], which is not consistent with the theory. However, the experimentally observed plateau switch in RhSi is messed up with non-topological contributions to the linear optical conductivity. Furthermore, the THz emission measurement used in this work reflects the transient current changing rate that is not exactly the current generation rate that is quantized. This indirect photocurrent measurement is strongly affected by the relaxation dynamics of photo-excited carriers. In addition, recent theoretical work including electron-electron interaction shows a correction on the quantization[260,261]. Thus, more studies are expected to further confirm this quantization effect.

## Perspectives

There are three important directions for further advancing the experimental progress of light-induced emerging phenomena: probing the intrinsic properties of the materials though optical response, classical light-driven dynamic tuning of the material properties through Floquet engineering, and quantum light-matter dressing by coupling to a cavity.

The coupling between light and spin or pseudospins, and the light-induced phenomena based on these couplings enable ample possibilities to probe and control intrinsic material properties by tuning light parameters. Novel pseudospins beyond sublattice, layer, valley, chirality and their couplings to spin and light are always interesting topics in 2D and topological materials. On the optical side, besides light energy and polarization, tuning knobs can also be phase, momentum, orbital angular momentum and pulse shape. Among these, the intricate coupling between orbital angular momentum and topological materials are largely unexplored, which may spark opportunities along this direction in the future[262]. Advanced laser technologies, such as precise phase and frequency control based on frequency combs[263] and precise manipulation of ultrafast pulse shaping[264], can further advance the light control capabilities. Compared to linear response, nonlinear optical response provides even richer light-induced phenomena. The geometric phase related enhancements on low order nonlinear optical response have already been demonstrated as a powerful experimental tool to probe topological properties of materials and further geometric phase related enhancement on higher order nonlinear responses remains to be confirmed experimentally. Unambiguous experimental verification of the authentic topological related nonlinear optical responses will benefit from future instrumental developments in the mid/far-infrared and terahertz wavelength range and versatile control of doping level through materials and devices engineering. We also expect to see more geometric phase related nonlinear optical response features, such as Chern number determined quantized response, to be proposed and verified[256].

As a classical way to couple the light field with matters, Floquet engineering provides exciting opportunities to change the dynamic properties of quantum materials, and furthermore, to tune their topological properties in non-equilibrium. However, there are challenges to achieve these goals from both experimental and theoretical points of view. Experimentally, the realizations of the abundant topological physics are still at early stages and substantial challenges need to be overcome. For example, ultrafast spectroscopies with stronger intensity and longer wavelength (MIR) are highly desired. Multiple pumping pulses have been added for extending the lifetime of the excitations[265,266]. Besides advanced experimental techniques, new quantum materials with high qualities are also important. Experimentally, the Floquet physics has been widely studied in cold atom[267] and photonic systems[268,269], however in solid-state materials where the coherence and dissipation are much more difficult to control, experimental observations of the Floquet physics have only been achieved for Bi$_2$Se$_3$[68], graphene[71], and WSe$_2$[124]. Reducing the inevitable heating and carrier scattering to prolong the lifetime of Floquet states is critical[124]. Along this line, several advanced theoretical methods beyond the Floquet theory have been developed to describe these driven-dissipative quantum systems, such as the non-equilibrium Keldysh-Green's function[270] that considers the dynamic evolution of a light-induced quantum state coupled with bathes[66,271]. Alternatively, the time-dependent density functional theory (TDDFT)[272,273] and quantum master equation could be used to calculate the dynamic evolution of the quantum materials under the optical driving. Via mapping the time-averaged wave function of the calculated steady state to a Floquet basis, information about the occupation of Floquet states can be provided directly, as has been demonstrated in Dirac semimetals[63,82] and graphene[72,74,274]. Meanwhile, other non-equilibrium theoretical approaches are also developed for the strongly correlated materials recently[275-281]. By combining both experimental and theoretical efforts, more experimental demonstrations of Floquet engineering as well as dynamic topological phase transitions in a wider range of materials are certainly exciting areas to look forward to in the near future.

In contrast to the classical Floquet theory, the hybrid light-matter states in the cavity could be described via the non-relativistic quantum electrodynamics (QED) formulation of light-matter dressing[282,283]. Interestingly, due to the quantum nature of the photon, the property of the quantum material could be controlled by tuning the nature of the quantum vacuum inside a resonant cavity even without external driving. Such technical pathway could avoid the influence of dissipation, decoherence and impurity scattering, which strongly limit the realization and survival of the Floquet state otherwise. Based on these concepts, the "cavity QED materials engineering" is proposed[180,284-287], which is similar to the Floquet engineering discussed in the second part of this review, and many interesting phenomena have been proposed, such as cavity enhanced ferroelectric[286,287], strong coupling in gated quantum dot[288], and strong nonlinearity of twisted van der Waals heterostructure inside a microcavity[289].

4

Acknowledgement:

This work is supported by the National Natural Science Foundation of China (Grant No. 11725418, 11427903, 12034001), National Key R & D Program of China (Grant No. 2016YFA0301004, 2020YFA0308800), Tsinghua University Initiative Scientific Research Program and Tohoku-Tsinghua Collaborative Research Fund, Beijing Advanced Innovation Center for Future Chip (ICFC), Beijing Nature Science Foundation (JQ19001).